\documentclass[useAMS,usenatbib]{mn2e}
\usepackage{xcolor}
\usepackage{amsmath, float}
\usepackage{txfonts}
\usepackage{fleqn}   
\usepackage{graphicx}
\usepackage{epstopdf}
\usepackage{verbatim}
\usepackage{mathtools}
\usepackage{bm}


\title[Jet-like features in BSPWNe]
{On the origin of jet-like features in bow shock pulsar wind nebulae}
\author[B. Olmi \& N. Bucciantini]{
B. Olmi$^{1,2,3}$ \thanks{E-mail: barbara@arcetri.astro.it} \& N. Bucciantini$^{1,4,5}$\\
$^{1}$INAF - Osservatorio Astrofisico di Arcetri, Largo E. Fermi 5,
I-50125 Firenze, Italy\\
$^{2}$Institute of Space Sciences (ICE, CSIC), Campus UAB, Carrer de Magrans s/n, 08193 Barcelona, Spain\\
$^{3}$Institut d'Estudis Espacials de Catalunya (IEEC), 08034 Barcelona, Spain\\
$^{4}$Dipartimento di Fisica e Astronomia, Universit\`a degli Studi di Firenze, Via G. Sansone 1, 
I-50019 Sesto F.~no  (Firenze), Italy\\
$^{5}$INFN - Sezione di Firenze, Via G. Sansone 1, I-50019 Sesto F.~no  (Firenze), Italy\\}

\begin{document}
 
\date{Accepted / Received}

\maketitle

\label{firstpage}

\begin{abstract}
Bow shock pulsar wind nebulae are a large class of non-thermal synchrotron sources associated to old pulsars, that have emerged from their parent supernova remnant and are directly interacting with the interstellar medium. Within this class a few objects show extended X-ray features, generally referred as ``jets'', that defies all the expectations from the canonical MHD models, being strongly misaligned respect to the pulsar direction of motion. It has been suggested that these jets might originate from high energy particles that escape from the system. Here we investigate this possibility, computing particle trajectories on top of a 3D relativistic MHD model of the flow and magnetic field structure, and we show not only that beamed escape is  possible, but that it can easily be asymmetric and charge separated, which as we will discuss are important aspects to explain known objects.
\end{abstract}


\section{Introduction}
\label{sec:intro}

Pulsar are generally born with a high kick velocity in the range 100-500
km s$^{-1}$  \citep{Cordes_Chernoff98a,Arzoumanian_Chernoff+02a,Sartore_Ripamonti+10a,Verbunt_Igoshev+17a}, and have lifetimes scales  \citep{Faucher-Giguere_Kaspi06a} much longer that the typical duration of their parent supernova remnants (SNRs). It is thus expected that the vast majority of them is found directly interacting with the interstellar medium (ISM). Bow shock pulsar wind nebulae (BSPWNe) form due to the confinement of the pulsar wind by the incoming (in the frame of the pulsar) ISM material, and indeed many of such systems have been observed in the past: some of them in optical H$_\alpha$ \citep{Kulkarni_Hester88a,Cordes_Romani+93a,Bell_Bailes+95a,van-Kerkwijk_Kulkarni01a,Jones_Stappers+02a,Brownsberger_Romani14a,Romani_Slane+17a},   others in non-thermal X-ray and radio \citep{Arzoumanian_Cordes+04a,Kargaltsev_Pavlov+17a,Kargaltsev_Misanovic+08a,Gaensler_van-der-Swaluw+04a,Yusef-Zadeh_Gaensler05a,Li_Lu+05a,Gaensler05a,Chatterjee_Gaensler+05a,Ng_Camilo+09a,Hales_Gaensler+09a,Ng_Gaensler+10a,De-Luca_Marelli+11a,Marelli_De-Luca+13a,Jakobsen_Tomsick+14a,Misanovic_Pavlov+08a,Posselt_Pavlov+17a,Klingler_Rangelov+16a,Ng_Bucciantini+12a}, in the UV \citep{Rangelov_Pavlov+16a} and IR
\citep{Wang_Kaplan+13a}. 
There are even cases of BSPWNe where the pulsar is moving through the parent supernova shell \citep{Frail_Giacani+96a, Swartz_Pavlov+15a, Temim_Slane+15a}.

Many of these systems show a typical cometary shape that conforms to the expectation of existing analytical and numerical models \citep{Bandiera93a,Wilkin96a,Bucciantini_Bandiera01a,Bucciantini02a, Bucciantini_Amato+05a, Olmi_Bucciantini19a,Barkov_Lyutikov+19b}. However, in few cases, non-thermal high energy emission is observed outside of the supposed location of the contact discontinuity separating the pulsar material from the ISM, where the canonical models predict there should be none. This emission ranges from faint X-ray haloes, as in the case of IC443 \citep{Olbert:2001, Bocchino:2001,Swartz_Pavlov+15a} and the Mouse \citep{Gaensler_van-der-Swaluw+04a}, or large non-thermal TeV haloes like in the case of Geminga \citep{Posselt_Pavlov+17a,Abeysekara_Albert+17a}, to more structured features like the X-ray {\it prongs} observed ahead of the bow shock in G327.1-1.1 \citep{Temim_Slane+15a}, or the one sided jets as seen in the Guitar Nebula \citep{Hui_Becker07a} or in the Lighthouse nebula  \citep{Pavan_Bordas+14a}.

It had been suggested \citep{Bandiera08a} that such peculiar features are due to the escape of particles at high energy (close to the pulsar voltage) that, once outside the bow shock, begin to stream along the ISM magnetic field, driving instabilities that amplifies it, and lead to particle confinement. However the mechanism leading to such escape, how it is related to the wind structure, and the geometry of the interaction with the ISM, how it depends on energy, and what causes the formation of one sided features, is still poorly understood. Reconnection has been invoked as a possible way to let particles from the pulsar wind flow in the ISM \citep{Barkov_Lyutikov+19a}. More recently \citet{Bucciantini:2018b}, has investigated the propagation of high energy particles in the magnetic field of a BSPWN, using a simplified, axisymmetric, laminar semi-analytic  model for the flow and magnetic field geometry. In that work it was found that a major role in ruling particles escape is played by current sheets and current layers, that can confine particles, producing strong asymmetries in their escape properties, and  leading also to charge separated outflows which are particularly relevant in the context of magnetic field amplification and self confinement.

Here we extend on that work by computing the transport of high energy particles, on top of a full high resolution 3D adaptive mesh refinement (AMR) relativistic MHD (RMHD) simulation of a BSPWN. Following  \citet{Bucciantini:2018b}, which suggested that the configurations more likely to produce asymmetric escapes where those with the relative inclination between the pulsar spin-axis and the ISM magnetic field different from either $0^\circ$ or $90^\circ$, and where the current sheets maintain their coherence, here we consider a case where the pulsar spin-axis is inclined by $45^\circ$ with respect to the pulsar kick velocity, while the magnetic field is inclined by $90^\circ$ with respect to the pulsar kick velocity and $45^\circ$ with respect to the pulsar spin-axis. 
The pulsar wind injection properties correspond to those of model $I_{\{\upi/4,\,0\}}$ in \citet{Olmi_Bucciantini19a} (from now on Paper I): an isotropic wind luminosity, and a high magnetization $\sigma =1$.
Unfortunately detailed information on the geometry of the pulsar wind - ISM interaction is not available for the few known systems showing these asymmetric features. However we have verified, considering some different configurations, that the one we are presenting here shows the highest level of anisotropy in the escape.
As clearly and amply discussed in \citet{Bandiera08a}, a full model of the emission and morphology of these features requires many more physical ingredients than simple MHD and particles orbits. In particular magnetic field turbulent amplification, in a transitional regime between free streaming and diffusion, together with synchrotron cooling, are mandatory to get the observed properties. Due to the complexity of   the dynamical scales involved, and the non-linearity of the processes at work, these are well beyond the possibility of the present numerical tools. Not to mention particles acceleration, which is still a matter of great debate. 

In  \citet{Bandiera08a} there were some fundamental questions left unanswered, in particular about the confinement of particles in the head of the bow shock, and on what was causing the feature to be asymmetric. In this work we attempted to answer those questions.

We also discuss a simple application of our results to the cases of the Guitar and Lighthouse nebulae.

The paper is organized as follows: in Sec.~\ref{sec:method}  we briefly describe the setup of our model and the method used to transport particles. In Sec.~\ref{sec:results} we show and discuss the results. In Sec.~\ref{sec:conclusions} we present our conclusions.

\section{Method}
\label{sec:method}

The magnetic field and flow structure is obtained from a numerical simulation, done following the same numerical setup of Paper I, to which the reader is referred for a more accurate description. The domain, in cartesian coordinates, is centered on the pulsar with the $z$-axis aligned with the pulsar kick velocity (the reference frame is moving with the pulsar, thus the ISM is seen as moving along the negative $z$ direction), and extends in the range [-17$d_0$, +17$d_0$] along the $x$ and $y$ directions and [-28$d_0$, 5$d_0$] along $z$, where $d_0$ is the so called {\it stand off distance} at which the wind momentum flux and the ISM ram pressure balance each other
\begin{align}
d_0=\sqrt{\dot{E}/(4\upi c \rho_{\rm ISM} v_{\rm PSR}^2})\,,
\end{align}
where $\dot{E}$ is the pulsar spin down luminosity, $\rho_{\rm ISM}$ is the ISM density, $v_{\rm PSR}$ the speed of the pulsar (PSR) with respect to the local medium, and $c$ the speed of light. The base grid have 128$^3$ equally spaced grid points. A set of four AMR levels is then used in order to reach the required resolution around the pulsar, and to resolve at best the bow shock head, which is a strongly dynamical region, corresponding to an effective resolution of 2048$^3$ cells at the highest level. The energy injection from the pulsar is taken to be isotropic (corresponding to $\alpha=0$ in Paper I), and the inclination of the pulsar spin-axis is $\phi_{\rm M}=45^\circ$. The wind is highly magnetized, with $\sigma =1$, leading to a more laminar structure with current sheets extending further in the tail. In this respect the simulation is equivalent to the case $I_{\{\upi/4,\,0\}}$ of Paper I. However, unlike in Paper I, a uniform magnetic field is added to the ISM aligned with the $y$-axis (laying in the same plane  of the pulsar kick velocity and spin-axis). The strength of the magnetic field is taken to be $B^2_{\rm ISM} = 0.01\rho_{\rm ISM} v_{\rm PSR}^2$.
For a typical ISM density of  $\rho_\mathrm{ISM} \simeq 1.6 \times10^{-24}$ g cm$^{-3}$ this is equivalent to  $\sim 3 \mu$G for a pulsar velocity of   $v_\mathrm{PSR} \simeq 2.5\times10^{7}$ cm s$^{-1}$, and $\sim 5 \mu$G for $v_\mathrm{PSR} \simeq 4\times10^{7}$ cm s$^{-1}$. 
For the {\it stand off distance}  we chose the value $d_0=5\times 10^{16}$ cm, typical of known BSPWNe.
The system is evolved in time until a quasi-stationary configuration is reached (usually after a few ISM crossing times).

Following \citet{Bucciantini:2018b}, given that we are interested in the escape  from
the very head of these nebulae, particles that move
backward at a distance exceeding $13d_0$ from the PSR are assumed to be lost in the tail. It is still likely that those particles
can escape from the tail, but given that the tail might be in general
more turbulent than the head, we expect the escape there to be more
isotropic. Moreover, we assume that all high energy particles
are injected from the PSR (or equivalently the pulsar wind) in the radial direction. The electric field is given by the ideal MHD condition $\boldsymbol{E}=-\boldsymbol{V}\times \boldsymbol{B}/c$ where $\boldsymbol{V}$ and $\boldsymbol{B}$ are the flow speed and magnetic field given by our RMHD simulation. All particles are injected with the same Lorentz factor, and we have investigated four cases with $\gamma=[0.5,1.0,3.0,10.]\times 10^7$, corresponding to values where we expect the transition from Larmor radii in the ISM magnetic field smaller that than $d_0$ to larger than $d_0$. Recall that what really matters for the escape is the magnetic rigidity, i.e. the ratio of the Larmor radius to the bow shock size, and that  results can be properly scaled \citep{Bucciantini:2018b}.

Particle trajectories in the electric and magnetic fields of this system are computed using an explicit Boris Pushing technique \citep{Boris:1972, Vay:2008, Higuera:2017}, which ensures energy and phase space number conservation in the non-radiative regime. We verified, by computing radiative losses, that they are always negligible. 
Interestingly, with the same algorithm we can evaluate the energy gain due to the electric field, and we found that it is also not important. In principle high energy particles injection might depend on the polar angle $\theta$ with respect to the pulsar spin-axis, and there is some evidence for this in  young PWNe \citep{Olmi_Del-Zanna+15a}. However in order to have enough statistic for all the injection angles, the same amount of particles, corresponding to about $2\times 10^5$, is injected every $10^\circ$ in polar angle (e.g. we inject the same amount of particles in the range $ \theta=$ [10,20]$^\circ$  as in $\theta=$ [80,90]$^\circ$). This allows us to sample with the same accuracy the fate of particles injected along the pulsar spin-axis as in the equatorial region. Particles are uniformly injected in time, and the system is evolved until a quasi-steady particles distribution is achieved (the particles injected at first reach the boundary of our domain).

\section{Results}
\label{sec:results}

We describe here what is the fate of particles injected from the PSR into the BSPWN. We follow the same approach that was used in \citet{Bucciantini:2018b}. For simplicity we only consider mono-energetic injection, such that we can evaluate how the probability of escape and  its properties depend on energy. We also consider particles of different sign $(-)$ and $(+)$, in order to model the charge dependence. Let us recall here that, depending on the relation between the sign of the charge and the polarity of the magnetic field, currents can play a confining or de-confining role, leading to charge separated flows. 

\begin{figure}
	\includegraphics[width=.45\textwidth,clip]{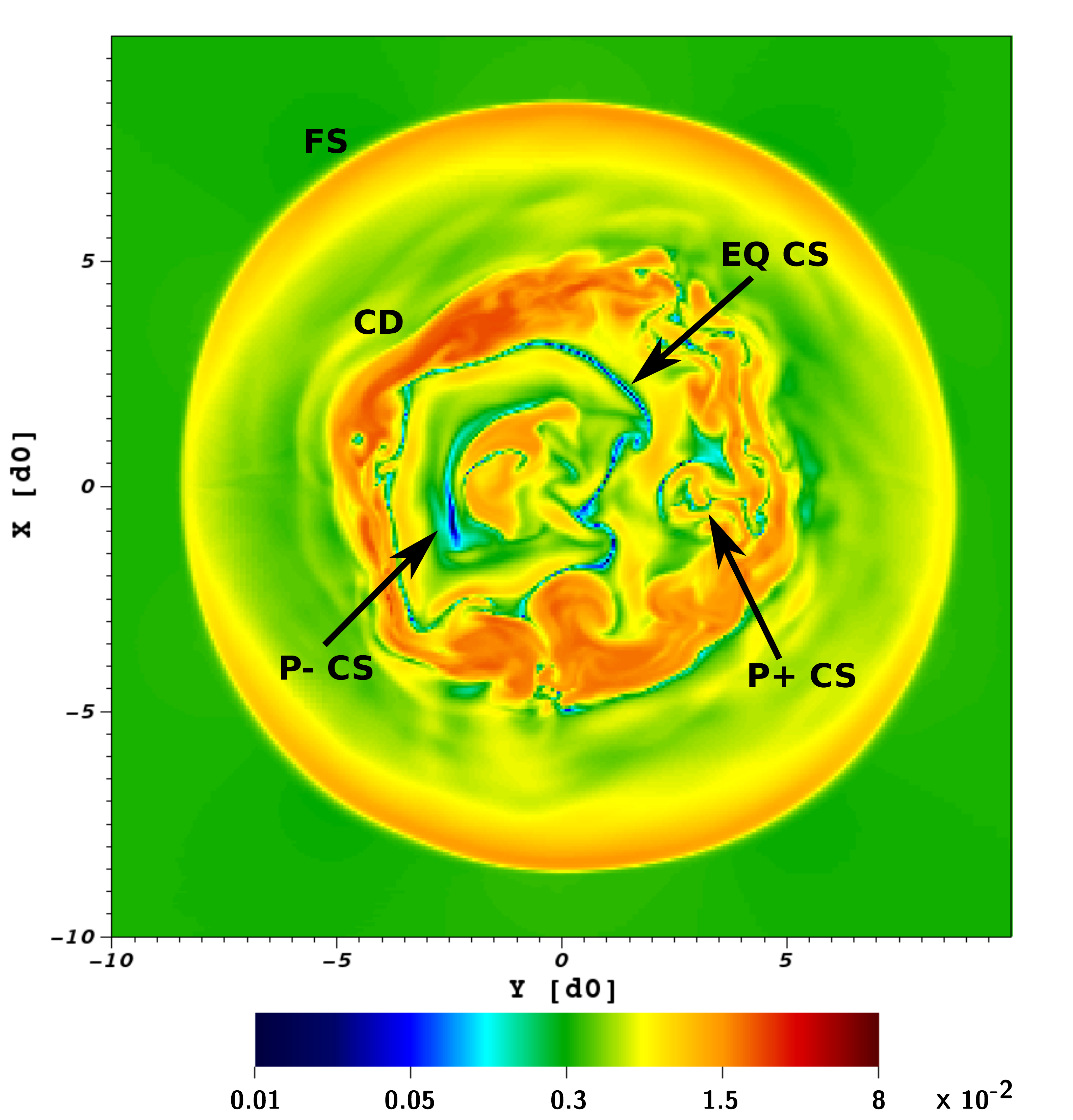}
	\caption{Transverse section of the BSPWN tail at $z=-11.5 d_0$. Color map of the intensity of the magnetic field in code units.  For the labels: ``FS'' indicates the position of  of the forward bow shock into the ISM; ``CD'' locates the position of the contact discontinuity between the ISM and PSR wind shocked components (the magnetopause); ``EQ CS'' traces the location of the equatorial current sheet; ``P-~CS'' locates the position of the polar current  coming from the PSR axis in the negative $y$-direction (the PSR axis that points away from the pulsar proper motion); ``P+~CS'' locates instead what is left of the polar current  coming from the PSR axis in the positive $y$-direction (the PSR axis that points toward the pulsar proper motion).
	The simulated data are read and elaborated mostly with the use of VisIt \citep{Childs:2012}, an opensource analysis tool. }
	\label{fig:1}
\end{figure}

In Fig.~\ref{fig:1} we show a cut of the BSPWN tail, where we highlight the position of the various current sheets and current lines, or what is left of them,  and the magnetopause at the contact discontinuity (CD) between the ISM material and the one coming from the PSR wind. It is evident that at the CD the PSR magnetic field is highly compressed, and this tends to create a magnetic barrier for low energy particles that prevent their escape into the ISM. It is also clear that the equatorial current sheet maintains its integrity in the tail (at least for the magnetic configuration of the BSPWN that we have selected, where turbulence is suppressed), and the same holds for the polar current associated to the PSR spin-axis pointing in the opposite direction with respect to the PSR motion (the one oriented in the negative $y$ and $z$ directions). This suggests that there is little or no chance for reconnection of the magnetic field lines in this region with the one of the ISM. On the other hand it is evident that the polar current originating from the PSR spin-axis pointing toward the PSR motion (the one oriented in the positive $y$ and $z$ directions) has been strongly affected by turbulence/reconnection and had lost its integrity. This immediately suggests that is it more likely for particles injected along this polar region $\theta =[0,60]^\circ$ to escape into the ISM.
%
The position in the tail where we show our results has been selected at $z=11.5d_0$ for two reasons.
We verified that most of the particles that escape do this in the very head ,within a few $d_0$ (at most 10) from the pulsar, and at $\sim 10 d_0$ all the particles that are confined in the tail tend to remain so.
On the other hand the resolution of our underlying simulation, due to the use of AMR, drops substantially beyond $12 d_0$. From that point numerical dissipation of the current sheets and layers becomes more important and this affects the propagation of particles, introducing numerical mixing and broadening of the tail (a similar effect of the AMR on the emissivity was noted in \citealt{Olmi:2019}).

\begin{figure*}
	\includegraphics[width=.95\textwidth,clip]{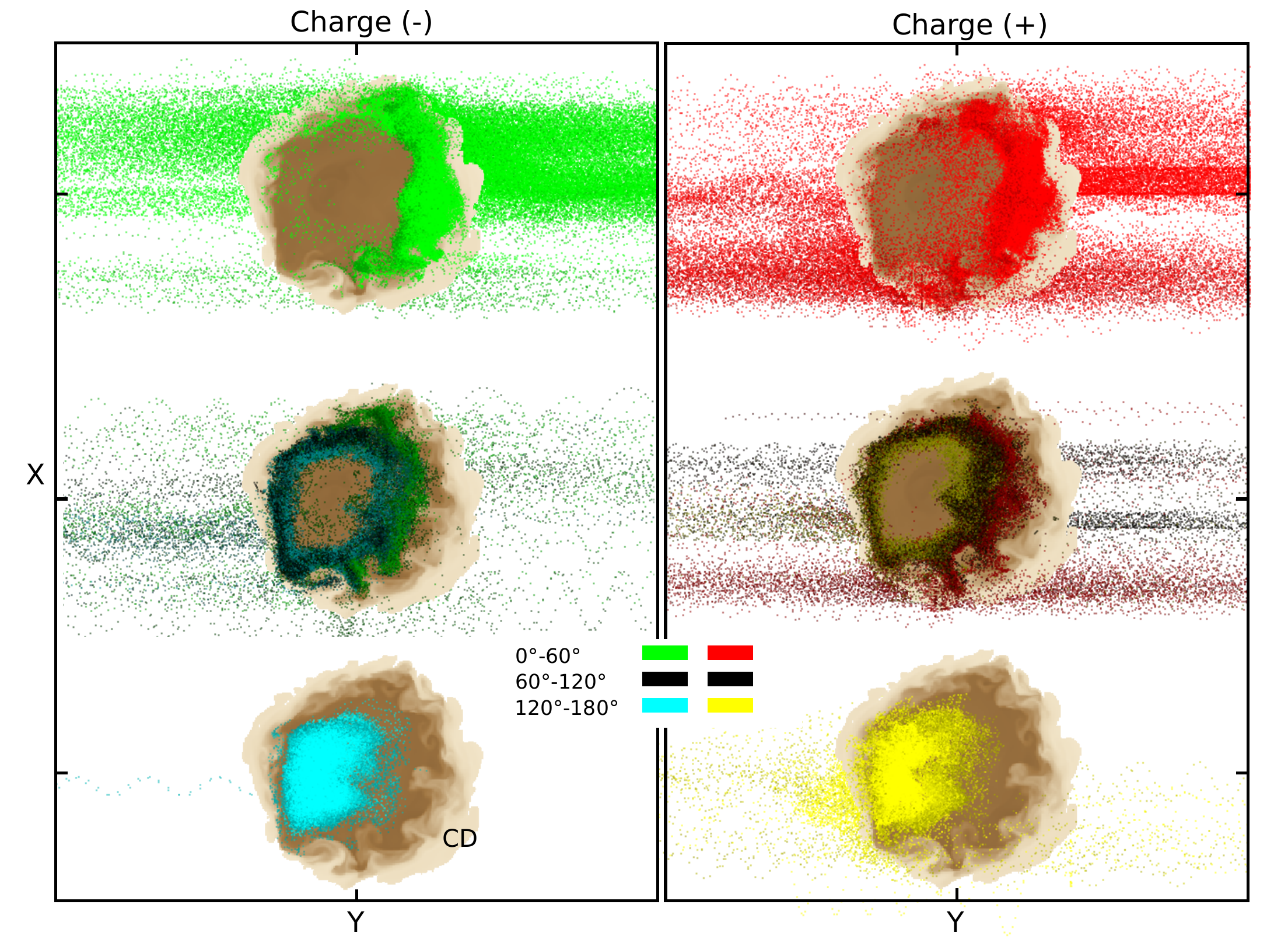}
	\caption{Projection on the $x-y$ plane of the 3D-positions of particles ($e^\pm$) injected in the wind with a Loretnz factor $\gamma=3\times 10^7$. Left and right panels refers to particles with different signs. From top to bottom, particles injected within different ranges of the polar angle $\theta$ with respect to the pulsar spin-axis: $[0,60]^\circ$ (green and red) indicates particles injected along the polar current originating from the PSR spin-axis pointing toward the PSR motion,   $[60,120]^\circ$ (black)  indicates particles injected along the equatorial current sheet,   $[120,180]^\circ$ (cyan  and yellow)  indicates particles injected along the polar current pointing in the opposite direction with respect to the PSR motion. The background image shows the $\log_{10}$  of the density (darker brown for lower values, lighter brown for higher ones), at $z=-11.5 d_0$ within the position of the CD, in order to mark the location of the shocked PSR wind. }
	\label{fig:2}
\end{figure*}

\begin{figure*}
	\includegraphics[width=.45\textwidth,clip]{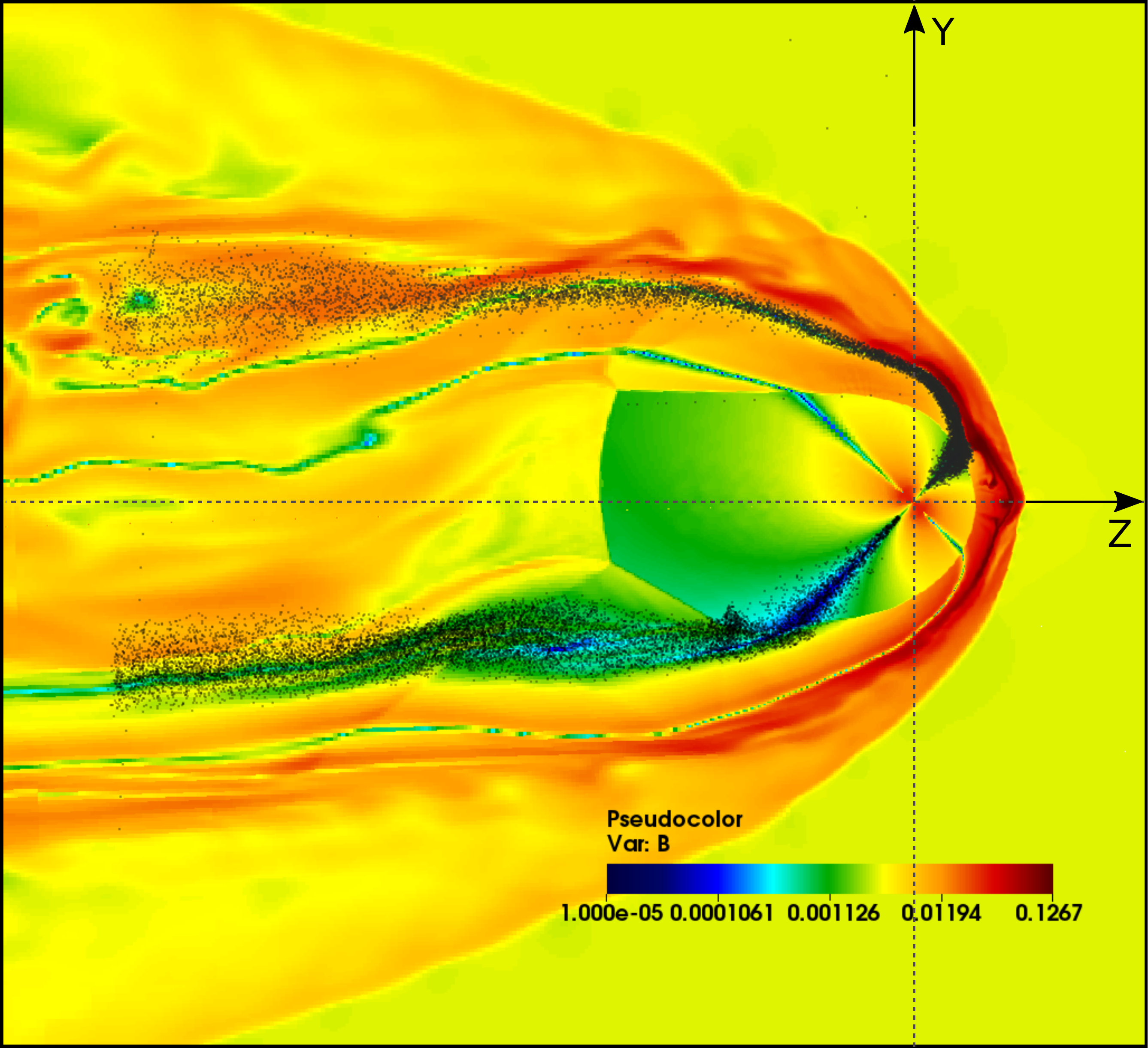}
	\includegraphics[width=.47\textwidth,clip]{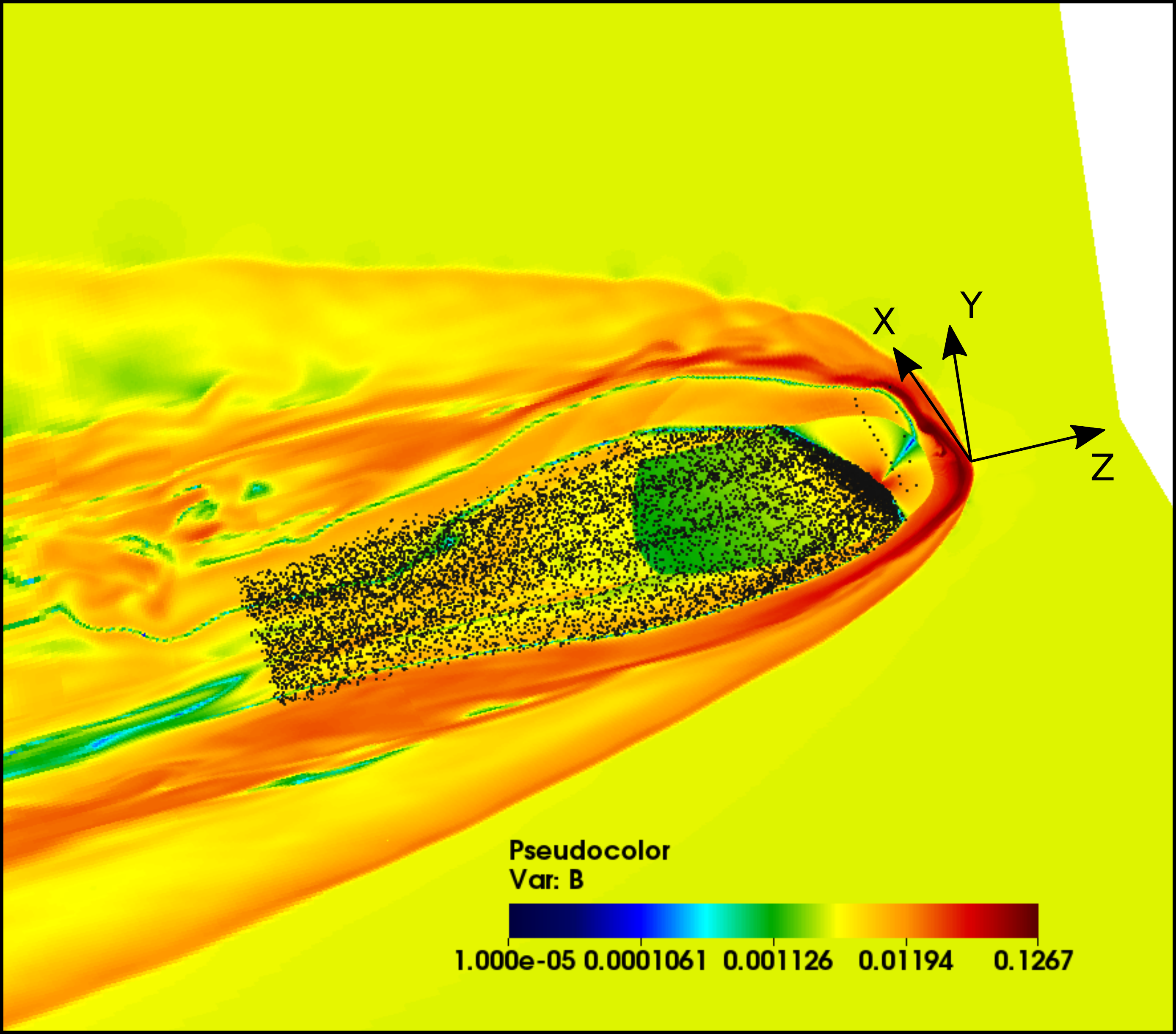}
	\caption{3D maps of selected particles superimposed on the 3D structure of the magnetic field in the BSPWN. On the left we show a 2D cut in the $y-z$ plane, at $x=0$ of the BSPWN color-coded according to the strength of the magnetic field in code units. Blue-cyan regions locate the position of the polar and equatorial currents. Over-imposed is the projection on the same plane of the position of particles injected in the polar regions $\theta = [0,10]^\circ$ and $\theta = [170,180]^\circ$, corresponding to the two PSR spin-axes pointing respectively toward and away from the PSR kick velocity (along positive $z$). On the right we show a 3D axonometric-projection of the BSPWN sliced along the $y-z$ half-plane and $x-z$, again color-coded according to the strength of the magnetic field in code units, together with the 3D positions of particles injected in the equatorial region $\theta = [85,95]^\circ$. }
	\label{fig:3}
\end{figure*}

In Fig.~\ref{fig:2} we show the projection on the  $x-y$ plane of the 3D-position of pairs injected with a Lorentz factor $\gamma= 3\times 10^7$, for both signs of their charge. Recall that what is relevant is not just the sign but its value with respect to the polarity of the magnetic field, as amply discussed and illustrated in \citet{Bucciantini:2018b}. It  is evident that particles injected with a polar angle $\ge 120^\circ$ (i.e. those injected around the polar current  relative to the PSR spin-axis pointing in the opposite direction with respect to the PSR motion), remain mostly confined in the tail within a region that is with good approximation delimited by the equatorial current sheet of Fig.~\ref{fig:1}. The same holds for particles injected  with a polar angle in the range  $[60,120]^\circ$, around the position of the equatorial current sheet. Those also show a clear tendency to remain confined between the equatorial current sheet and the magnetopause at the CD, even if now there is a larger fraction of them that manage to escape into the ISM. 
On the contrary, those particles injected with a polar angle $\le 60^\circ$ (i.e. those injected around the polar current  associated  to the PSR spin-axis pointing toward the PSR motion) have a high chance of escaping in the ISM, with just some of them residually confined close to the position labelled ``P+~CS''  in Fig.~\ref{fig:1}.
 This is a clear evidence that the magnetic field lines associated to this polar current tend to reconnect/interact with the ISM magnetic field, most likely because of turbulence at the CD.
  This makes easier for particles injected in that polar region to jump on ISM magnetic field lines and finally escape in the unperturbed ISM. 
  This is most clearly shown in Fig.~\ref{fig:3}, where the 3D position of particles in the BSPWN is shown superimposed on a map of the magnetic field to highlight the correlation between the particles position and the current sheets and lines.

What is also evident from Fig.~\ref{fig:2} is that the escape is strongly charge dependent, and non-symmetric. For example the $(-)$ particles  injected with a polar angle $\le 60^\circ$, show a clear asymmetry toward the positive $y$-direction, where the pulsar spin-axis points. This is less marked for the $(-)$ particles, most likely because of different confinement properties of the polar currents with respect to the sign of particles, as discussed in \citet{Bucciantini:2018b}.  
The trend seems reversed at  $\theta \ge 120^\circ$, where some of the $(+)$ particles manage to escape, unlike the $(-)$ ones. 
It is also evident that, at this energy, escaping particles remain confined into a ISM magnetic flux tube with typical dimension of $\sim 10 d_0$, with no sign of further diffusion (in part due to the lack of magnetic turbulence in the ISM of our simulation). 
It is also interesting to notice that the escape seems to take place within narrow $\sim d_0$ streams, likely associated to small reconnection regions or turbulent patches. 
It is indeed clear that these streams connect the ISM to those positions along the CD where strong turbulence develops, and the integrity of the polar current line is broken. 
It is also evident that these streams are charge separated, in the sense that there are streams dominated by $(+)$ particles and other by  $(-)$ ones. Most evident is the up-down (along the $x$-direction) asymmetry of the  particles  injected with a polar angle $\le 60^\circ$ (green and red dots in Fig.~\ref{fig:2}).

\begin{figure*}
	\includegraphics[width=.95\textwidth,clip]{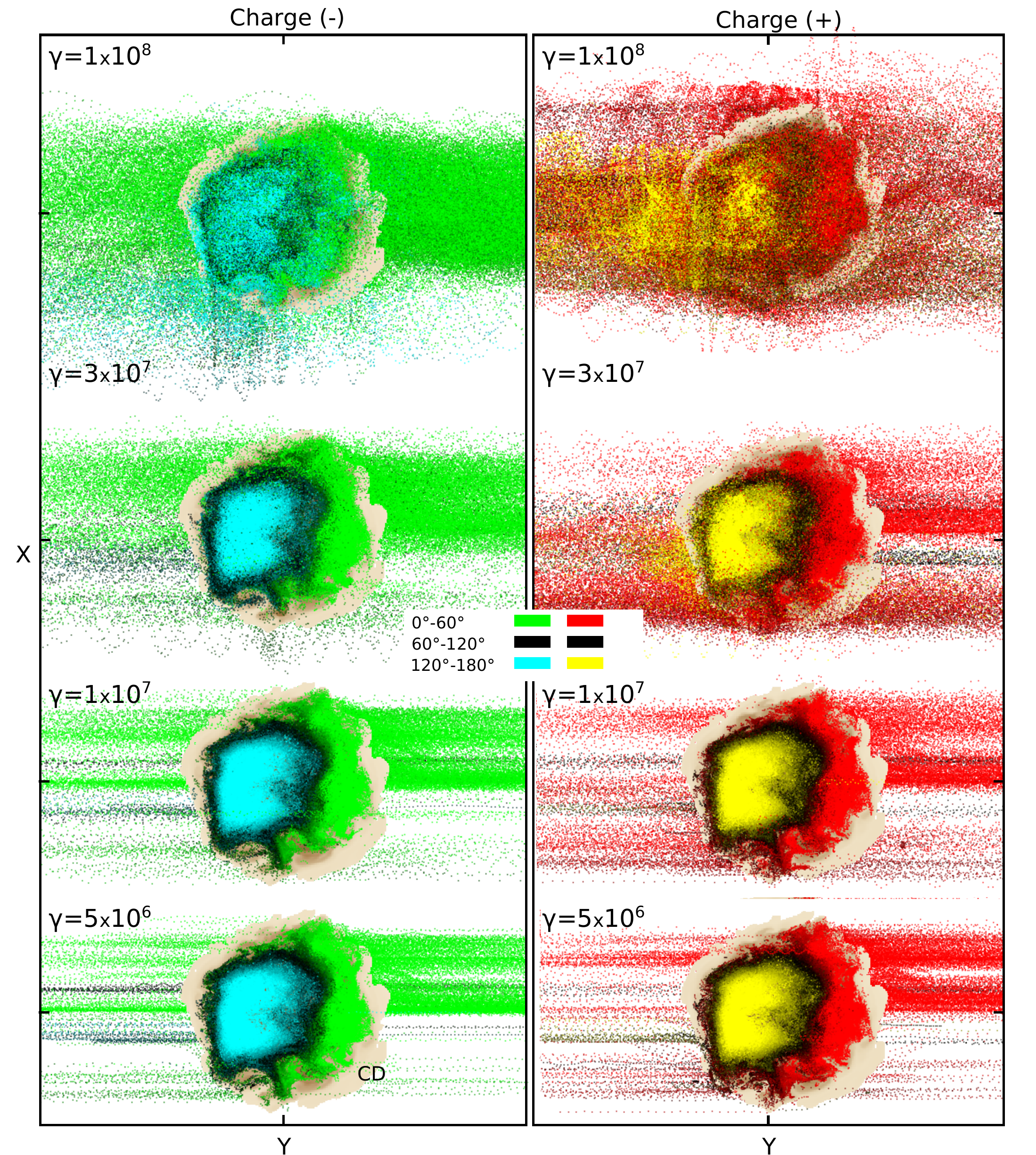}
	\caption{Projection on the $x-y$ plane of the 3D-positions of particles ($e^\pm$) injected in the wind with various Lorenz factors. From bottom to top:  $\gamma= [0.5,1,3,10]\times 10^7$. Left and right panels refers to particles with different signs color-coded according to the range of the polar angle $\theta$ of injection (as in previous Fig.\ref{fig:2}). The background image shows the $\log_{10}$  of the density (darker brown for lower values, lighter brown for higher ones), at $z=-11.5 d_0$ within the position of the CD, in order to mark the location of the shocked PSR wind. }
	\label{fig:4}
\end{figure*}

\begin{figure}
	\includegraphics[width=.5\textwidth,clip]{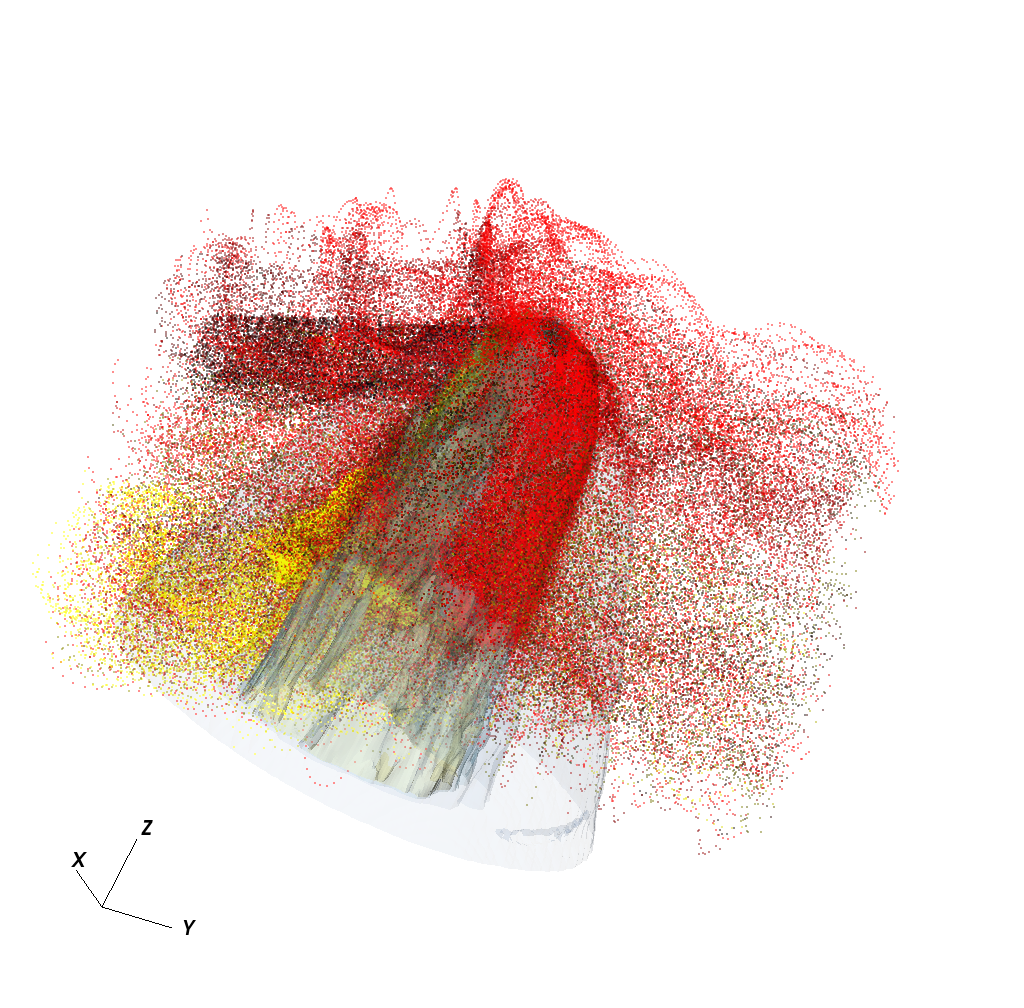}
	\caption{3D maps of selected particles superimposed on the 3D structure of the BSPWN, for $(+)$ particles with  $\gamma= 10^8$, color-coded according to their injection angle $\theta$, as in Fig.~\ref{fig:4}}
	\label{fig:5}
\end{figure}

We have also investigated how much the properties of the escape depend on the energy of the particles. In Fig.~\ref{fig:4} we show the projections on the  $x-y$ plane of the 3D-positions of pairs injected with Lorentz factors $\gamma= [0.5,1,3,10]\times 10^7$, for both signs of the charge and color-coded according to their polar injection angle $\theta$. As one would naively expect, at low energies particles follow the flow of matter, and in fact only a very marginal fraction of them escapes into the ISM, and almost exclusively those injected within a range $\theta = [0,60]^\circ$. 
This seems to agree with the idea that those particles are tightly bound to magnetic field lines around which they gyrate, and that only where these lines reconnect with the ones of the ISM, they manage to escape.  Again there is a clear evidence that particles tend to escape in charge separated streams that are not symmetric, and at low energies there is a clear preference for the escape along the positive $y$-direction, that is where the front PSR spin-axis points.  At $\gamma= 3\times 10^7$ we begin to see a transition: the number of escaping particles increases, the presence of streams is less evident, there is a tendency for particles to be more diffuse, and the escape tends to become more symmetric, yet the confining role of current sheet is still present. We are in a regime where particles cannot be considered tied to magnetic field lines: their Larmor radii begin comparable with the size and thickness of the current  sheets, so that they can easily jump from a field line to another, especially where turbulence tends to mix them. 
At  $\gamma= 10^8$, particles begin to diffuse efficiently outside the BSPWN: there is less evidence for any confinement in the tail, where particles injected in different regions tend to mix with each other, and even particles injected at  $\theta = [120,180]^\circ$ manage to escape in the ISM. Now particles looks more diffuse, with some tendency for the $(+)$ ones to be less confined than the $(-)$ ones. There is little evidence for an asymmetry in the $y$-direction. In 3D some particles, in particular those injected in a range $\theta = [0,60]^\circ$, are also found ahead of the bow-shock as shown in Fig.~\ref{fig:5}. This marks the transition to the diffusive regime: now particles have Larmor radii, that even in the strong magnetic field compressed at the magnetopause, are of the order of the BSPWN thickness (the distance between the termination shock of the PSR wind  and the CD).

\subsection{The Lighthouse Nebula}
\label{subsec:light}
The Lighthouse nebula, associated with PSR J$1101-6101$, shows an X-ray feature extending for $\sim 11 \mathcal{D}_p$ pc and with a X-ray luminosity of $L_X\simeq 4\times 10^{33} \mathcal{D}_p^2$ erg/s, where $\mathcal{D}_p$ is the distance in units of 7 kpc \citep{Pavan_Bordas+14a}.
The pulsar has an estimated spin-down luminosity of $\dot{E}\simeq 1.36\times 10^{36}$ erg/s \citep{Halpern:2014} and moves through a medium with an estimated density of $0.1$ particles/cm$^{-3}$ \citep{Tomsick:2012} and a velocity of $\simeq 10^8$ cm/s \citep{Pavan_Bordas+14a}.
Assuming equipartition with a magnetic field of the order of $15 \; \mu$G \citep{Pavan_Bordas+14a} we can estimate what fraction of the pulsar energy needs to be supplied through escaping particles in order to ensure the total inferred energetics. This requires the knowledge of the thickness of the feature along the line of sight, which cannot be derived from observations, but can be inferred from our results to be of the order of a few $d_0$ (the width-thickness of the escaping stream).
We find that, to power the energetics of the X-ray jet, the escaping particles must carry a fraction of the order of $\eta_J=3\times 10^{-3} \mathcal{D}_p $ of the pulsar spin-down power. This corresponds to a fraction of the order of $10^{-2}$ of the energy injected within $60^\circ$ from the polar axis by an isotropic wind, and it is fully compatible with our findings.
Note moreover that the fraction $\eta_J$ equals the ratio $L_X/\dot{E}$, suggesting that most of the energy is rapidly radiated away.

\subsection{The Guitar Nebula}
\label{subsec:guitar}
The Guitar nebula, associated with PSR B2224+65, shows an X-ray feature extending for $\sim 1 \mathcal{D}_p$ pc, characterized by an X-ray luminosity of $L_X\simeq 4.1\times 10^{31} \mathcal{D}_p^2$ erg/s, where $\mathcal{D}_p$ is now in units of 1.86 kpc \citep{Bandiera08a,Hui_Becker07a}.
The pulsar has an estimated spin-down luminosity of $\dot{E}\simeq 1.2\times 10^{33}$ erg/s and moves through a medium with an estimated density of $3\times10^{-2} \mathcal{D}_p^{-4}$ particles/cm$^{-3}$ and a velocity of $\simeq1.6\times  10^8 \mathcal{D}_p$ cm/s \citep{Chatterjee:2002}.
Proceeding as before and assuming equipartition with a magnetic field of $20 \, \mu$G (the half of the maximum value estimated by \citet{Bandiera08a} based on synchrotron cooling), we find that to power the energetics of the X-ray jet, the escaping particles must now carry a fraction of the order of $\eta_J=3\times 10^{-2} \mathcal{D}_p^3 $ of the pulsar spin-down power (note that the different dependence on the distance $\mathcal{D}_p$ is due to the fact that the ISM density is here inferred from  the bow-shock size and not independently estimated as for the Lighthouse nebula).
This corresponds to a fraction of the order of $10^{-1}$ of the energy injected within $60^\circ$ from the polar axis by an isotropic wind, and it is again compatible with our findings, in terms of the fraction of the escaping particles.
Here again we find that the fraction $\eta_J$ equals the ratio $L_X/\dot{E}$.

\section{Conclusions}
\label{sec:conclusions}

In this work we have extended our previous results on the escape of high energy particles from BSPWNe, by computing particle trajectories on top of a realistic 3D flow structure from RMHD simulations. Given the complexity of the problem, and the size of the parameters space in terms of the pulsar wind properties and ISM magnetic field configurations, we limited the present analysis to a selected configuration, where the relative inclinations among the pulsar spin-axis, the pulsar kick velocity and the ISM magnetic field are expected to maximize the degree of asymmetry in the outflow.  

We show that as the energy increases, there is a transition in the properties of escaping particles, and that in general are only those coming from the frontal polar region of the pulsar wind that manage to escape efficiently, while the others remain confined in the tail. At low energy the escape seems to be associated to reconnection between magnetic field lines of the pulsar wind and ISM, in the sense that particles gyrating along those magnetic field lines can move from the PWN to the ISM, as suggested by \citet{Barkov_Lyutikov+19a}, and indeed there is a small charge asymmetry, less than $10\%$. 
The outflows are also strongly asymmetric (by a factor 4 to 5), reflecting likely a similar property of the reconnection at the magnetopause, and limited to particles injected in the frontal polar region of the pulsar wind, $\theta =[0,60]^\circ$. 
The fact that, in this low energy regime, the escaping streams are small and separated is consistent with the fact that reconnection at the magnetopause is known to be patchy and sporadic, and to lead only to marginal flux transfer \citep{Kan88a,Pinnock_Rodger+95a,Fear_Trenchi+17a}. 

As the energy increases however the escape enters a different regime. 
Outflows become more charge separated, indicating that the confinement role played by current sheets and current lines enters into the picture, as suggested in  \citet{Bucciantini:2018b}. 
The charge asymmetry is now about a factor 1.5, while the spatial asymmetry is a factor 2. 
Finally, one enters a fully diffusive regime \citep{Bykov:2017}, where the charge asymmetry exceeds a factor 2, while the spatial asymmetry is reduced to less than 1.5. 
Interestingly, for the parameters of our model, this transition takes place within just an order of magnitude in the energy of the particles from $\gamma=10^7$ to $\gamma=10^8$. This suggests that the escaping flow should be quasi-monochromatic. 

The asymmetry in the escaping flux, as well as the fact that they are charge separated, can explain the presence of one-sided jets, as in the Guitar and Lighthouse nebulae. 
We recall here that the relation between the number of escaping particles and the presence of bright features is non trivial: in fact \citet{Bandiera08a} has shown that some form of magnetic field amplification and particle confinement, more likely driven by turbulence injected by the same escaping particles, is necessary to explain the observed luminosity. 
Even a factor of two in the level of asymmetry could be important.  
Moreover, in the presence of a net current, the magnetic field can be
amplified much more efficiently than for a neutral flow \citep{Skilling71a,Bell04a}. 
This means that self confinement is more efficient. 
However, here we have shown that the energetics of these jets are fully compatible with the expected particles escape of our numerical models.
Our results also show what could be a possible explanation for the X-ray morphology of G327.1-1.1 \citep{Temim_Slane+15a}. 
The SNR shell is known to be subject to strong turbulence and instability that can mix the magnetic field and produce field lines aligned in the radial direction \citep{Chevalier:1992, Chevalier:1995, Blondin_Chevalier+01a, Dubner_Giacani15a}. The {\it prongs} could be the equivalent of the streams we see in our simulations, but now oriented along some of these radial field lines, while the diffuse emission could be due to high energy particles. 
The fact that in this system it is still possible to identify a bow shock could be due to lower energy particles that manage to remain confined in the tail.

Our results suggest that the two long tails seen in Geminga \citep{Posselt_Pavlov+17a}, that cannot be explained as due to limb brightening, could be the result of high energy particles injected along the polar axis of the pulsar and confined along the polar current, bent backward by the interaction with the ISM. This is similar to what is seen in the left panel of Fig.~\ref{fig:3}, but in the case of an orthogonal configuration $\phi_M=90^\circ$.  

We plan in the future to extend this study considering a larger set of configurations, including cases where the magnetic field is not orthogonal to the pulsar proper motion, that could potentially lead to an even stronger level of asymmetry or more turbulent configurations, where the escape of particles might be diffusive even at low energies. 
Our conclusion is that the existence of asymmetric features points toward a magnetic field configuration, in terms of relative inclinations, close to the one we have selected. This is in principle a falsifiable prediction if, in future, independent measures of the pulsar wind magnetic field structure and of the ISM magnetic field were to be available.

More important, in the present study we assumed the background MHD model to be time independent. This is marginally reasonable for models where the flow tends to stay laminar, even if the typical timescale for particles to escape are comparable or only slightly longer than the flow time in the shocked PSR wind. One would need to compute particle trajectories on a time dependent model, which at this moment is not feasible with the AMR code we have been using (PLUTO, \citealt{Mignone:2012}), and requires the development of new numerical tools. 

Finally we recall, as discussed in \citet{Bucciantini:2018b}, that there are several other possible sources of asymmetry in the escape of particles from BSPWNe: from wind anisotropy, to differential acceleration at the termination shock. However our results show that even if isotropy is assumed in the injection, asymmetric escape is a common outcome.

\section*{Acknowledgements}
We acknowledge the ``Accordo Quadro INAF-CINECA (2017-2019)''  for the availability of high performance computing resources and support. Simulations have been performed as part of the class-A project ``Escape of particles from bow shock pulsar wind nebulae'' (PI B. Olmi). 
 The authors also acknowledge financial support from the ``Accordo Attuativo ASI-INAF n. 2017-14-H.0 Progetto: \textit{on the escape of cosmic rays and their impact on the background plasma}'' and from the INFN Teongrav collaboration.
As usual, the authors acknowledge Andrea Mignone for fundamental code support. The authors wish also to acknowledge the anonymous Referee for her/his useful comments, that helps in improving this manuscript.
\bibliographystyle{mn2e}
\bibliography{Bib}

\end{document}